\documentclass[traditabstract]{aa}
\usepackage{amssymb, amsmath}
\usepackage[below]{placeins}
\usepackage{graphicx,natbib,txfonts}
\usepackage{verbatim}
\usepackage{subfigure}
\usepackage{color}

\begin{document}
\title{Joint reconstruction of galaxy clusters from gravitational lensing and thermal gas}
\subtitle{II. Inversion of the thermal Sunyaev-Zel'dovich effect}
\author
 {Charles L. Majer\thanks{\email{C.Majer@stud.uni-heidelberg.de}} \and Sven Meyer \and Sara Konrad \and Eleonora Sarli \and Matthias Bartelmann}
\institute
 {Universit\"at Heidelberg, Zentrum f\"ur Astronomie, Institut f\"ur Theoretische Astrophysik, Philosophenweg 12, 69120 Heidelberg, Germany}
\date{\today}

\abstract
 {This paper continues a series in which we intend to show how all observables of galaxy clusters can be combined to recover the two-dimensional, projected gravitational potential of individual clusters. Our goal is to develop a non-parametric algorithm for joint cluster reconstruction taking all cluster observables into account. In this paper, we begin with the relation between the Compton-$y$ parameter and the Newtonian gravitational potential, assuming hydrostatic equilibrium and a polytropic stratification of the intracluster gas. We show how Richardson-Lucy deconvolution can be used to convert the intensity change of the CMB due to the thermal Sunyaev-Zel'dovich effect into an estimate for the two-dimensional gravitational potential. Synthetic data simulated with characteristics of the ALMA telescope show that the two-dimensional potential of a cluster with mass $5\times10^{14}\,h^{-1}M_\odot$ at redshift $0.2$ is possible with an error of $\lesssim5\,\%$ between the cluster centre and a radius $r\lesssim0.9\,h^{-1}\,\mathrm{Mpc}$.}

\keywords{(Cosmology:) dark matter, Galaxies: clusters: general, Gravitational lensing: strong, Gravitational lensing: weak}

\maketitle

\section{Introduction}

Beginning with \citep[][hereafter NFW]{NA97.1, NA96.1}, a multitude of numerical simulations has shown that gravitationally bound structures dominated by dark matter should exhibit a universal density profile with three characteristic properties: it starts flat in the core, steepens around a scale radius $r_\mathrm{s}$, and asymptotically approaches a double-logarithmic slope near $-3$ towards the virial radius $r_\mathrm{vir}$ (e.g.\ \citealt{JI00.1, ME06.1, NA04.1, PO03.1, MO98.3, MO99.2}; see also \citealt{EI89.1}). The concentration parameter $c = r_\mathrm{vir}/r_\mathrm{s}$ is found in simulations to depend only weakly on the mass $M$. For cold dark matter, it decreases $\propto M^{-0.1}$ \citep[see][for examples]{NA97.1, NA96.1, BU01.1, DO04.2, DU08.1, EK01.1, GA08.1, MA08.2, MA07.5, NE07.1, SE00.1, SH06.2, ZH09.1}

Moreover, cold dark matter is expected to clump on virtually all scales. Massive objects such as galaxy clusters should thus have a broad spectrum of massive sublumps embedded \citep[cf.][]{BO09.1, DO09.2, GA11.1, GA04.2, GI10.2, ZE05.1}

While these mostly qualitative statements hold for gravitationally bound structures composed of cold dark matter on all mass scales, the dark-matter distribution in galaxy clusters should be least affected by baryonic physics because in them, the cooling time of the baryonic matter almost everywhere exceeds the Hubble time. Galaxy clusters are thus perhaps the class of objects best suited for testing whether the expectations raised by simulations are indeed supported by real dark-matter structures.

Galaxy clusters provide a multitude of observables. Weak and strong gravitational lensing effects allow direct inversions yielding the scaled surface-mass density or, equivalently, the effective lensing potential of a lensing mass distribution. X-ray emission and the thermal Sunyaev-Zel'dovich (hereafter SZ) effect reflect the physical state of the hot intracluster gas. Assuming equilibrium and stability, the gas properties can also be related to the gravitational potential. This suggests to devise a method by which lensing, X-ray, and thermal SZ data can be combined in a joint analysis aiming at recovering the gravitational potential best compatible with all observables.

The advantage and at the same time the main obstacle in such an approach is that the observables probe the gravitational potential on different scales. Strong lensing typically occurs on angular scales smaller than $\approx(20-30)''$, which at the same time approximately equals the resolution limit of weak lensing. Strong and weak lensing can be combined in the same analysis either by assuming a parametric form of the density profile to be adapted to both effects, or in a parameter-free approach fitting the effective lensing potential on a grid whose resolution is adapted to the angular scales on which the effects occur. Based on a maximum-likelihood cluster-reconstruction method \citep{BA96.3}, we have in the past years developed a multi-scale approach to parameter-free reconstructions of the lensing potential which has been shown to perform reliably with simulated data \citep{CA06.1, ME09.2, ME10.2}, and which has been applied to several clusters so far. The method adapts potential values in grid cells until the lensing data are reproduced optimally in a least-$\chi^2$ sense \citep[see also][]{BR05.2, BR06.1}.

The series of papers which the present paper belongs to aims at extending this method towards including all available cluster observables in a joint reconstruction aiming at the one gravitational potential underlying all of them. In the first paper \citep{PI}, we have shown how X-ray data can be analysed to recover the gravitational potential projected along the line-of-sight. Assuming hydrostatic equilibrium and a polytropic gas stratification, we could show that the projected gravitational potential can be reproduced with typical errors at the per-cent level under typical conditions. For simplicity, not for necessity, we have further assumed spherical symmetry there. Analysed this way, the X-ray data can now be combined with the lensing data by adding a suitable term to the $\chi^2$ function initally defined for lensing.

This paper describes how this approach can be extended towards including thermal SZ data. Again for simplicity, we adopt a spherical galaxy-cluster model for simulating observations with the signal and noise characteristics of ALMA.

We begin in Sect.~2 with a brief overview of the thermal SZ effect and review the essential steps of our method. We keep this short and refer the reader to the detailed description in the first paper of this series where possible. In Sect.~3, we describe how we simulated thermal SZ data sets including a realistic noise model. We test our algorithm varying its input parameters and estimate the error of the reconstruction in Sect.~4. We estimate the quality of our reconstruction of the lensing potential by bootstrapping 200 realizations of one specific example of a galaxy cluster. In Sect.~5, we draw our conclusions.

\section{Recovering the projected gravitational potential from SZ data}

\subsection{The thermal Sunyaev-Zel'dovich effect}

The thermal Sunyaev-Zel'dovich effect (SZ) \citep{Sunyaev1980} is caused the hot electrons in the intracluster plasma that inversely Compton scatter the much less energetic photons of the cosmic microwave background (CMB) to higher energies. The SZ effect thus slightly distorts the CMB spectrum away from its Planckian shape. Seen against the CMB, clusters cast shadows below 217~GHz and shine at frequencies above. Massive galaxy clusters can be identified by this characteristic spectral appearance at centimetre to millimetre wavelengths if the angular resolution of the telescope is of order $1'$ or better, for example with the South Pole Telescope \citep{Ruhl2004}.

The thermal SZ effect is quantified by the Compton-$y$ parameter
\begin{equation}
  y(\vec s) = \frac{k_\mathrm{B}}{m_\mathrm{e}c^2}\sigma_\mathrm{T}
  \int\mathrm{d} z\,T(\vec s, z)\rho_\mathrm{e}(\vec s, z)\;,
\label{eq:3}
\end{equation} 
which compares the mean thermal energy $k_\mathrm{B}T$ with the rest energy $m_\mathrm{e}c^2$ of the electron and multiplies their ratio with the Thomson cross section $\sigma_\mathrm{T}$ and the electron number density $\rho_\mathrm{e}$. By Compton-upscattering, the specific intensity of the CMB seen through a galaxy cluster changes by
\begin{equation}
  \frac{\Delta I_\mathrm{SZ}(\vec s)}{B_\omega(T)} = g(x)y(\vec s)\;,
\label{eq:4}
\end{equation}
relative to the Planck spectrum $B_\omega(T)$ of the CMB, where
\begin{equation}
  x = \frac{\hbar\omega}{k_\mathrm{B}T}
\end{equation} 
is the photon energy in units of the mean thermal energy, and
\begin{equation}
  g(x) = \frac{x\mathrm{e}^x}{\mathrm{e}^x-1}\left(x\frac{\mathrm{e}^x+1}{\mathrm{e}^x-1}-4\right)\;,
\end{equation}
describes the frequency dependence of the thermal SZ effect.

Due to the finite telescope resolution, the observable quantity is not quite the Compton-$y$ parameter or the specific CMB intensity change from Eq. (\ref{eq:4}), but rather the beam-convolved intensity change or Compton-$y$ parameter. The Compton-$y$ profile convolved with a beam profile $b(\vec s)$ is
\begin{equation}
  \bar y(\vec s) = \int\mathrm{d}^2s'\,y(\vec s')b(\vec s-\vec s')\;.
\label{eq:5}
\end{equation}

\subsection{Basic relations}

An ideal gas assumed to have a polytropic stratification with index $\gamma$ attains the density
\begin{equation}
  \rho = \rho_0\varphi^{1/(\gamma-1)}\;,
\label{eq:6}
\end{equation} 
in hydrostatic equilibrium with the suitably scaled, dimension-less gravitational potential
\begin{equation}
  \varphi = \frac{\gamma-1}{\gamma}\frac{\rho_0}{P_0}\left(\Phi_\mathrm{cut}-\Phi\right)\;.
\end{equation}
Here, $\Phi$ is the Newtonian gravitational potential, $\Phi_\mathrm{cut}$ is the potential value where the density of the bound gas is supposed to drop to zero, and $\rho_0$ and $P_0$ are fiducial values for the gas density and its pressure at an arbitrary, suitably chosen radius. By the ideal gas equation, the gas temperature is
\begin{equation}
  T = T_0\varphi\;,
\label{eq:7}
\end{equation}
also relative to the temperature $T_0$ at a fiducial radius, for which we choose the virial radius without loss of generality. Equations (\ref{eq:6}) and (\ref{eq:8}) were derived in detail in KMMSB.

Combining Equations (\ref{eq:6}), (\ref{eq:7}) and (\ref{eq:3}), the Compton-$y$ parameter can be rewritten in terms of the gravitational potential $\varphi$ as
\begin{equation}
  y(\vec s) = \frac{k_\mathrm{B}\sigma_\mathrm{T}}{m_\mathrm{e}c^2}T_0\rho_0
  \int\mathrm{d} z\,\varphi^\eta(\vec s, z)\;,
\label{eq:8}
\end{equation}
where the exponent $\eta = \gamma(\gamma-1)^{-1}$. For realistic polytropic indices, $1.1\lesssim\gamma\lesssim1.2$ \citep{Finoguenov2001}, the exponent $\eta$ is quite a large number, $6\lesssim\eta\lesssim11$.

Using Eq.~(\ref{eq:5}), the beam-convolved Compton-$y$ parameter can be re-written as
\begin{align}
  \bar y(\vec s) &= \int\mathrm{d}^2s'\,y(\vec s')b(\vec s-\vec s')\nonumber\\&=
  \frac{k_\mathrm{B}\sigma_\mathrm{T}}{m_\mathrm{e}c^2}T_0\rho_0
  \int\mathrm{d} z\int\mathrm{d}^2s'\,\varphi^\eta(\vec s', z)\,b(\vec s-\vec s')\label{eq:9}\\&=
  \frac{k_\mathrm{B}\sigma_\mathrm{T}}{m_\mathrm{e}c^2}T_0\rho_0
  \int\mathrm{d} z\,\overline{\varphi^\eta}(\vec s, z)\;,\nonumber
\end{align}
where the overbar abbreviates the convolution.

Combining the last expression with Eq.~(\ref{eq:4}) leads to a result which can be identified with an effective, three-dimensional, projected pressure $P(\vec r)$,
\begin{align}
  \frac{\Delta\bar I_\mathrm{SZ}(\vec s)}{B_\omega(T)} &=
  g(x)\frac{k_\mathrm{B}\sigma_\mathrm{T}}{m_\mathrm{e}c^2}T_0\rho_0
  \int\mathrm{d} z\,\overline{\varphi^\eta}(\vec s, z)\nonumber\\&=
  \int\mathrm{d} z\,P(\vec s, z)\;.
\label{eq:11}
\end{align}

We shall argue now that the beam convolution can in fact be ignored for our purposes. Current and future SZ observations reach an angular resolution very much better than the angular resolution that can be achieved with potential reconstructions based on gravitational lensing. Assuming 50 background galaxies per square arc minute and averaging over 20 galaxies to obtain a sufficiently robust weak-lensing signal, the resolution of a weak-lensing map corresponds to $\approx35''$. Beam profiles of modern thermal-SZ observations are much narrower than that, allowing us to approximate the beam $b$ in (\ref{eq:11}) by a Dirac delta distribution. This paper uses a configuration of ALMA which results in a beam size of $1''$ and is thus larger than the resolution of strong lensing observations (e.g. HST/ACS with $\approx0.1''$).

From Eq.~(\ref{eq:11}), we can then infer the relation
\begin{equation}
  P(\vec r) =  P_0\varphi^\eta(\vec s, z)\;,
\label{eq:12}
\end{equation}
between the effective pressure $P(\vec r)$ and the scaled gravitational potential $\varphi$, where the amplitude
\begin{equation}
  P_0 = g(x)\frac{k_\mathrm{B}\sigma_\mathrm{T}}{m_\mathrm{e}c^2}T_0\rho_0\;,
\end{equation}
was introduced.

Equation (\ref{eq:12}) serves as the basis for a method analogous to that presented in KMMSB for recovering the projected gravitational potential:

\begin{enumerate}
  \item By deprojection of the measured, relative specific intensity change $\Delta I_\mathrm{SZ}B^{-1}_\omega(T)$, we find an estimate for the three-dimensional effective pressure $P$. As described in KMMSB, this can be achieved by means of the Richardson-Lucy algorithm.
  \item Next, we use Eq.~(\ref{eq:12}) to find an estimate $\tilde\varphi$ for the scaled gravitational potential $\varphi$,
  \begin{equation}
    \tilde\varphi = \left(\frac{P(\vec r)}{P_0}\right)^{1/\eta}\;.
  \label{eq:13}
  \end{equation}
  \item The estimate $\tilde\varphi$ of the three-dimensional potential is then projected to find an estimate $\tilde\psi$ for the two-dimensional potential,
  \begin{equation}
    \tilde\psi(\vec s) = \int\mathrm{d} z\,\tilde\varphi(\vec s, z)\;.
  \label{eq:14}
  \end{equation}
\end{enumerate}

Since $\eta$ is large, the exponent $1/\eta$ is a small number, which is a most welcome property of Eq.~(\ref{eq:12}). Fluctuations in the estimate $\tilde P$ of the deprojected effective pressure will be substantially smoothed that way.

\subsection{Deprojection}

As described in detail in KMMSB, we adopt the Richardson-Lucy algorithm \citep{Lucy1974, Lucy1994} for deprojection. For simplicity, we assume spherical symmetry for now, which is not a necessary requirement and will be relaxed in further work.

Generally, Richardson-Lucy deprojection connects two functions, say $f(r)$ and $g(s)$, related by the projection
\begin{align}
  g(s) &= \int\mathrm{d} r\,K(s|r)f(r)\;,\\
  f(r) &= \int\mathrm{d} s\,K'(r|s)g(s)\;,
\label{eq:1516}
\end{align}
mediated by the projection kernel $K(s|r)$ and the deprojection kernel $K'(r|s)$. In our application, $f(r)$ represents a function given in three dimensions and $g(s)$ its projection onto the sky. For a spherically-symmetric body, the normalised projection kernel $K(s|r)$ is
\begin{equation}
  K(s|r) = \frac{2r}{\pi}\frac{\Theta(r^2-s^2)}{\sqrt{r^2-s^2}}\;,
\label{eq:17}
\end{equation}
where $\Theta(x)$ is the Heaviside step function. The projection and deprojection kernels are related by Bayes' theorem,
\begin{equation}
  K'(r|s) = \frac{f(r)}{g(s)}\,K(s|r)\;.
\end{equation} 

Since $f(r)$ is unknown, so is the deprojection kernel $K'(r|s)$. However, given an estimate $\tilde f_i(r)$ for the function $f(r)$, an estimate $\tilde g_i(s)$ of the projection $g(s)$ is
\begin{equation}
  \tilde g_i(s) = \int\mathrm{d} r\,K(s|r)\tilde f_i(r)\;,
\label{eq:18}
\end{equation} 
implying the estimate
\begin{equation}
  \tilde K'(r|s) = \frac{\tilde f_i(r)}{\tilde g_i(s)}\,K(s|r)\;,
\label{eq:19}
\end{equation}
for the deprojection kernel. This suggests an iterative scheme: Given an estimate $\tilde f_i(r)$ for the function $f(r)$ at the iteration level $i$, an improved estimate $\tilde f_{i+1}(r)$ is found by
\begin{equation}
  \tilde f_{i+1}(r) = \tilde f_i(r)\int\frac{g(s)}{\tilde g_i(s)}\,K(s|r)\;.
\label{eq:19a}
\end{equation} 

Including a regularisation term for suppressing small-scale fluctuations, the change in the estimate $\tilde f_i(r)$ by the $i$-th iteration step is
\begin{equation}
  \Delta\tilde f_i = \Delta_H\tilde f_i+\Delta_S\tilde f_i\;,
\label{eq:20}
\end{equation}
where $\Delta_H\tilde f_i(r)$ represents the change $\tilde f_{i+1}-\tilde f_i$ given by Eq.~(\ref{eq:19a}), and $\Delta_S\tilde f_i$ may contain an entropic term $S$ against a suitably chosen prior $\chi$.

Specialising the Richardson-Lucy algorithm to our specific case, we replace $g(s)$ by the observable intensity change $\Delta I_\mathrm{SZ}B_\omega^{-1}(T)$ relative to the Planck spectrum of the CMB and $f(r)$ by the effective pressure $P(r)$.

Identifying the projected function $g(s)$ with the radial profile of specific intensity change $\Delta I_{\mathrm{SZE}}(s)$ and the deprojected function $f(r)$ with the effective pressure $P(r)$, the complete iteration including an entropic regularisation term $S$ reads
\begin{equation}
  \Delta\tilde P_i = \tilde P_i\left[
    \int\mathrm{d} s\,\frac{\Delta I_\mathrm{SZ}(s)}{\Delta\tilde I_{\mathrm{SZ}, i}(s)}\,K(s|r)-
    1-\alpha\left(\ln\frac{\Delta\tilde I_{\mathrm{SZ}, i}(s)}{\chi}+S\right)
  \right]\;,
\label{eq:21}
\end{equation}
with
\begin{equation}
  \Delta\tilde I_{\mathrm{SZ}, i}({s}) = \int\mathrm{d} r\,K(s|r)\tilde P_i(r)\;.
\end{equation}

The complete algorithm is characterised by two parameters, the amplitude of the regularisation term $\alpha$ and the smoothing-scale length $L$, contained in the prior $\chi$. The value $\alpha = 0$ and $L = 0\,h^{-1}\,\mathrm{Mpc}$ correspond to no regularisation and no smoothing, respectively.

\begin{figure}[!ht]
  \includegraphics[width=\hsize]{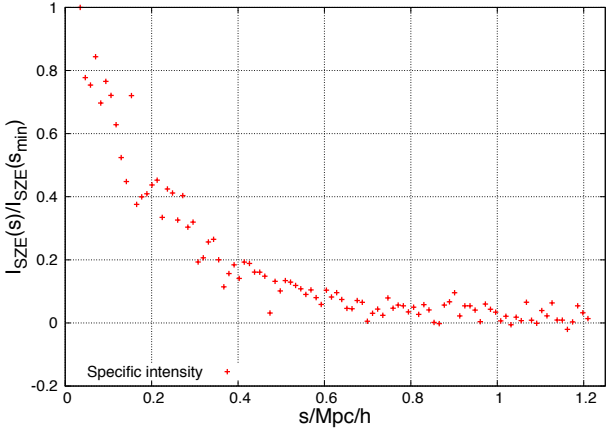}
\caption{Azimuthally averaged and normalized specific intensity change profile of a simulated galaxy cluster with mass $5\times10^{14}\,h^{-1}M_\odot$ at redshift $0.2$. The displayed maximal range of radial values corresponds to the virial radius of the cluster.}
\label{fig:cluster_profile}
\end{figure}

\section{Simulating SZ observations}

\subsection{Simulating SZ observations}

For testing the algorithm sketched above, we simulate thermal SZ signal of a massive galaxy cluster, assuming spherical symmetry, hydrostatic equilibrium and an NFW density profile $\rho(r)$ 
\begin{equation}
  \rho(r) = \frac{\rho_\mathrm{s}}{(r/r_\mathrm{s})(1+r/r_\mathrm{s})^2}\;,
\label{eq:22}
\end{equation}
with the scale radius $r_\mathrm{s} = r_{200}/c$ and the characteristic density $\rho_\mathrm{s}$ of the halo. We choose the concentration parameter $c = 5$ and a spatially flat standard $\Lambda$CDM cosmology with $\Omega_\mathrm{m0} = 0.3$, $\Omega_\mathrm{b0} = 0.04$ and $\Omega_{\Lambda0} = 0.7$. 
The properties of the ICM are chosen as follows:

\begin{itemize}
  \item The plasma contains 75\% hydrogen and 25\% helium by mass. Both components are completely ionised.
  \item The gas-mass fraction equals the universal baryon mass fraction $f_\mathrm{b} = \Omega_\mathrm{b}/\Omega_\mathrm{m}$.
  \item The gas has a constant polytropic index of $\gamma = 1.2$.
\end{itemize}

The gas density and temperature profiles are then calculated using Eqs.~(\ref{eq:6}) and (\ref{eq:7}). To obtain a temperature profile which drops to zero at a large radius, we choose a large cut-off radius for the gravitational potential of $r_\mathrm{cut} = 100\,r_{200}$. Given the density and temperature profiles, the specific intensity change $\Delta I_\mathrm{SZ}$ can be calculated using Eq.~(\ref{eq:12}).

The scale radius of our simulated cluster is $r_\mathrm{s} = 0.25\,h^{-1}\mathrm{Mpc}$, its virial radius is $r_\mathrm{vir} = 1.25\,h^{-1}\mathrm{Mpc}$.

\subsection{Background fluctuations due to unresolved clusters}

Galaxy clusters unresolved by the telescope beam contribute a background noise level $y_\mathrm{bg}$ that needs to be taken into account in all following calculations.

Clusters are unresolved if they appear (much) smaller than the beam size. The background signal is thus dominated by low-mass clusters \citep{Bartelmann2001}. Since an ideally homogeneous background could be removed from the data, we only have to consider the average background fluctuation level $\Delta y_\mathrm{bg}$. The mean background level contributed by unresolved clusters is
\begin{align}
  y_\mathrm{bg} &= \int\mathrm{d} z\left|\frac{\mathrm{d} V}{\mathrm{d} z}\right|(1+z)^3
  \int\mathrm{d} M\,n(M, z)Y(M, z)\nonumber\\&=
  \int\mathrm{d} M\int\mathrm{d} V\,Y(M, z)\frac{\mathrm{d}^2N(M, z)}{\mathrm{d} M\mathrm{d} V}\;,
\label{eq:23}
\end{align}
where the mass- and redshift integrations have to be carried out over that area in the mass-redshift plane where clusters are unresolved. The integrated Compton-$y$ parameter $Y(M,z)$ of a cluster with mass $M$ at redshift $z$ is
\begin{equation}
  Y = \int\mathrm{d}\vec s\,y(\vec s) = \frac{k_\mathrm{B}T}{m_\mathrm{e}c^2}
  \frac{\sigma_\mathrm{T}}{D_\mathrm{d}^2}N_\mathrm{e}\;,
\label{eq:24}
\end{equation}
where the angular-diameter distance $D_\mathrm{d}$ to the cluster appears. The total number of (hot) electrons is $N_\mathrm{e}$.

We choose the cluster mass function $n(M,z)$ described by the Sheth-Tormen model \citep{Sheth1999},
\begin{align}
  n(M, z)\,\mathrm{d} M = A\sqrt{\frac{2}{\pi}}\,\left(
    1+\frac{1}{\nu^{2q}}
  \right)\frac{\bar\rho}{M}\frac{\mathrm{d}\nu}{\mathrm{d} M}
  \exp\left(-\frac{\nu^2}{2}\right)\,M\;,
\end{align}
where $\nu=\sqrt{a}\delta_{c}[\sigma_{0}(M)D_+(z)]^{-1}$ is the linear amplitude required for the collapse of a density fluctuation with the present \textit{rms} fluctuation $\sigma_0(M)$. The linear growth factor of the density perturbations is $D_+$ and the critical linear density contrast for non-linear collapse is $\delta_\mathrm{c}$. $\bar\rho$ is the mean background density at the present epoch. For the remaining parameters, \citet{Sheth2001} find $A = 0.322$, $a = 0.707$ and $q = 0.3$.

The number of background sources is a discrete probability distribution and therefore follows a Poissonian distribution. But due to the very high number of background galaxy clusters the Poissonian distribution, according to the central limit theorem, converges pointwise towards the normal distribution.
Therefore background thermal-SZ fluctuations can be assumed to be given by a Gaussian fluctuation of the number of clusters per unit mass and volume, if one also neglects any cluster correlations. Thus, the rms background fluctuation is
\begin{equation}
  \Delta y_\mathrm{bg} = \left[
    \int\mathrm{d} M\int\mathrm{d} V\,Y^2(M, z)\frac{\mathrm{d}^2N(M, z)}{\mathrm{d} M\mathrm{d} V}
  \right]^{1/2}\;.
\label{eq:26}
\end{equation}
Following Eq.~(\ref{eq:26}), each point in the cluster coordinate frame follows a Gaussian distribution with a mean equal to the sum of the analytic value of the Compton-$y$ parameter and the signal of the unresolved clusters. The variance of the distribution is then equal to the rms of the background fluctuations according to  Eq.~(\ref{eq:26}).

\subsection{Instrument noise}

Another relevant source of noise is the measurement noise of the telescope and the detector. We assume that the error has a Gaussian distribution with standard deviation $\sigma$ around zero, with $\sigma$ depending on the telescope and detector configuration and the setup of the specific observation.

Since we need a resolution for our simulation that is comparable to other observations of massive galaxy clusters, we choose the Atacama Large Millimeter/submillimeter Array (ALMA) as an example telescope for achieving sufficiently precise thermal-SZ observations. For all simulations presented in the remainder of this paper, we use the following configuration of ALMA:

All $N = 32$ antennae are assumed to be available with a baseline of $650\,\mathrm{m}$. This configuration has an angular resolution of $1.0''$ in a frequency band chosen to be centred on $\nu = 116\,\mathrm{GHz}$. Its bandwidth is assumed to be $\Delta\nu = 7.5\,\mathrm{GHz}$.

According to the ALMA user manual \citep{ALMA} the point-source sensitivity $\sigma$ of ALMA in units of Jansky is
\begin{equation}
  \sigma = \frac{2k_\mathrm{B}T_\mathrm{sys}}
  {\eta_\mathrm{q}\eta_\mathrm{c}A_\mathrm{eff}\sqrt{N(N-1)n_\mathrm{p}\Delta\nu t_\mathrm{int}}}\;,
\label{eq:32}
\end{equation}
where $T_\mathrm{sys}$ is the temperature of the system, $\eta_\mathrm{q}$ the quantum efficiency (e.g.\ $\eta_\mathrm{q} = 0.96$), $\eta_\mathrm{c}$ the correlator efficiency (e.g.\ $\eta_\mathrm{c} = 0.88$), $A_\mathrm{eff}$ the effective area of the antenna, $n_\mathrm{p}$ the number of polarisation states (e.g.\ $n_\mathrm{p} = 2$) and $t_\mathrm{int}$ is the integration time. Given a required sensitivity, we choose the integration time to be 5 days. Given the standard deviation of the instrument noise, the final observed image can be simulated assuming a Gaussian distribution of the signal with the variance $\sigma^2$.

For this configuration, the radial profile of the thermal SZ signal can be obtained by averaging the simulated image in circles around the cluster centre. This profile is shown in Fig.~\ref{fig:cluster_profile} for one realisation of the simulated galaxy cluster. At a radius near $0.8\,h^{-1}\mathrm{Mpc}$ (i.e.\ $3.5r_\mathrm{s}$), the cluster signal sinks below the instrumental noise and the background-fluctuation level.

\begin{figure}[!ht]
  \includegraphics[width=\hsize]{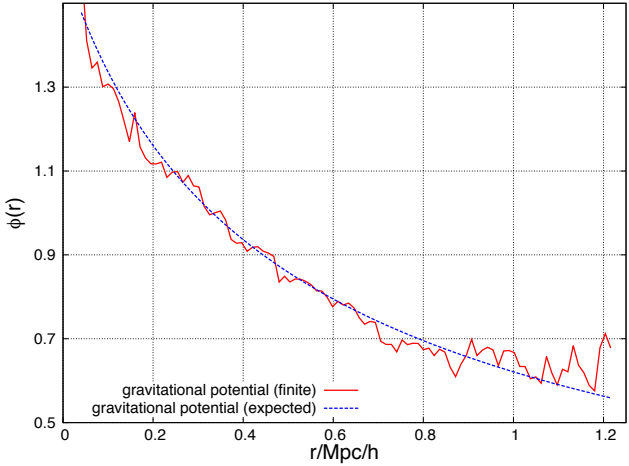}
\caption{Reconstructed scaled gravitational potential $\varphi$ of a simulated galaxy cluster with mass $5\times10^{14}\,h^{-1}M_\odot$ at redshift $0.2$. The potential was reconstructed using $\alpha = 0.2$ and $L = 0.3\,h^{-1}\,\mathrm{Mpc}$ in the regularisation term.}
\label{fig:gravpot}
\end{figure}

\begin{figure*}[!ht]
  \subfigure[]{\includegraphics[width=0.48\hsize]{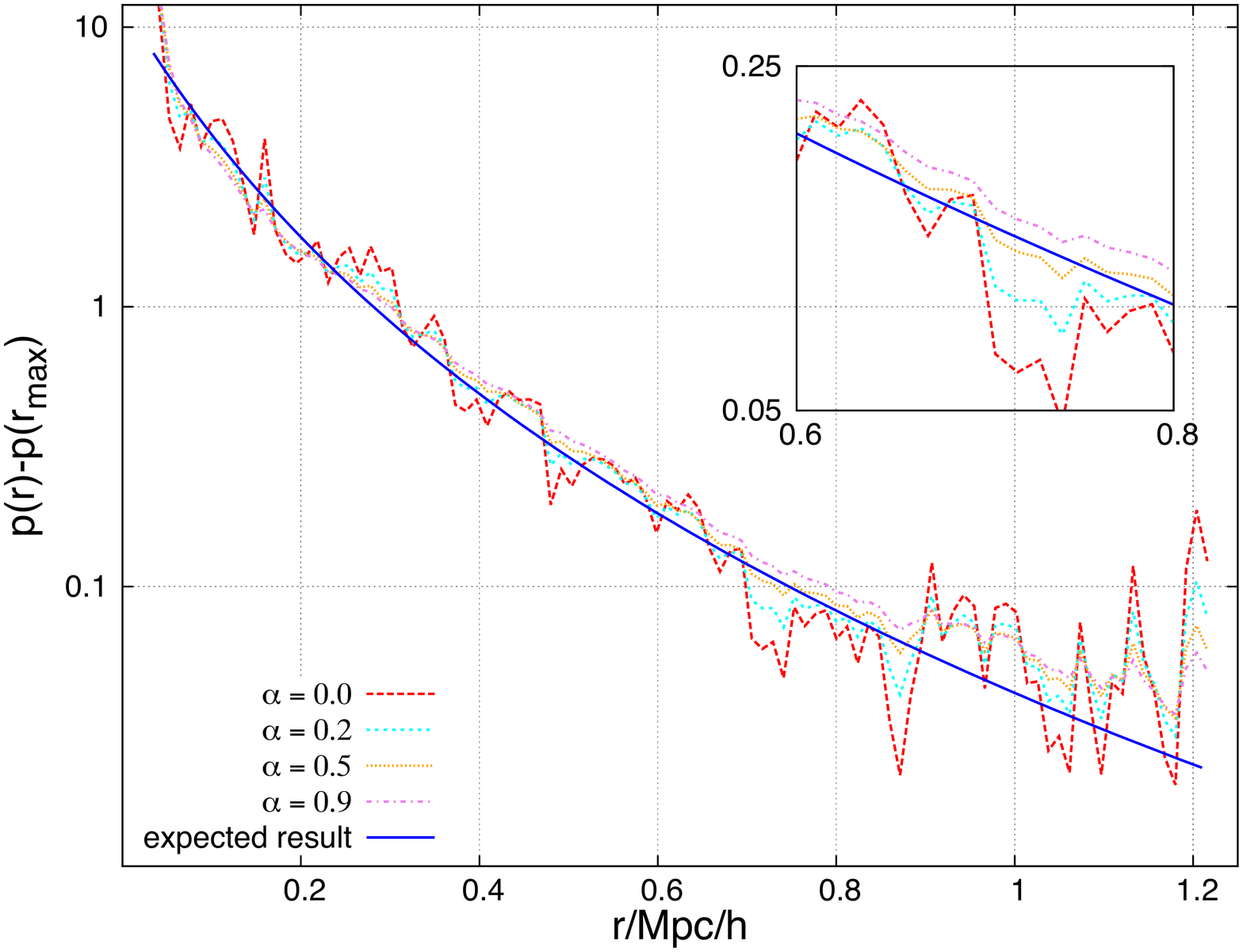}}\hfill
  \subfigure[]{\includegraphics[width=0.48\hsize]{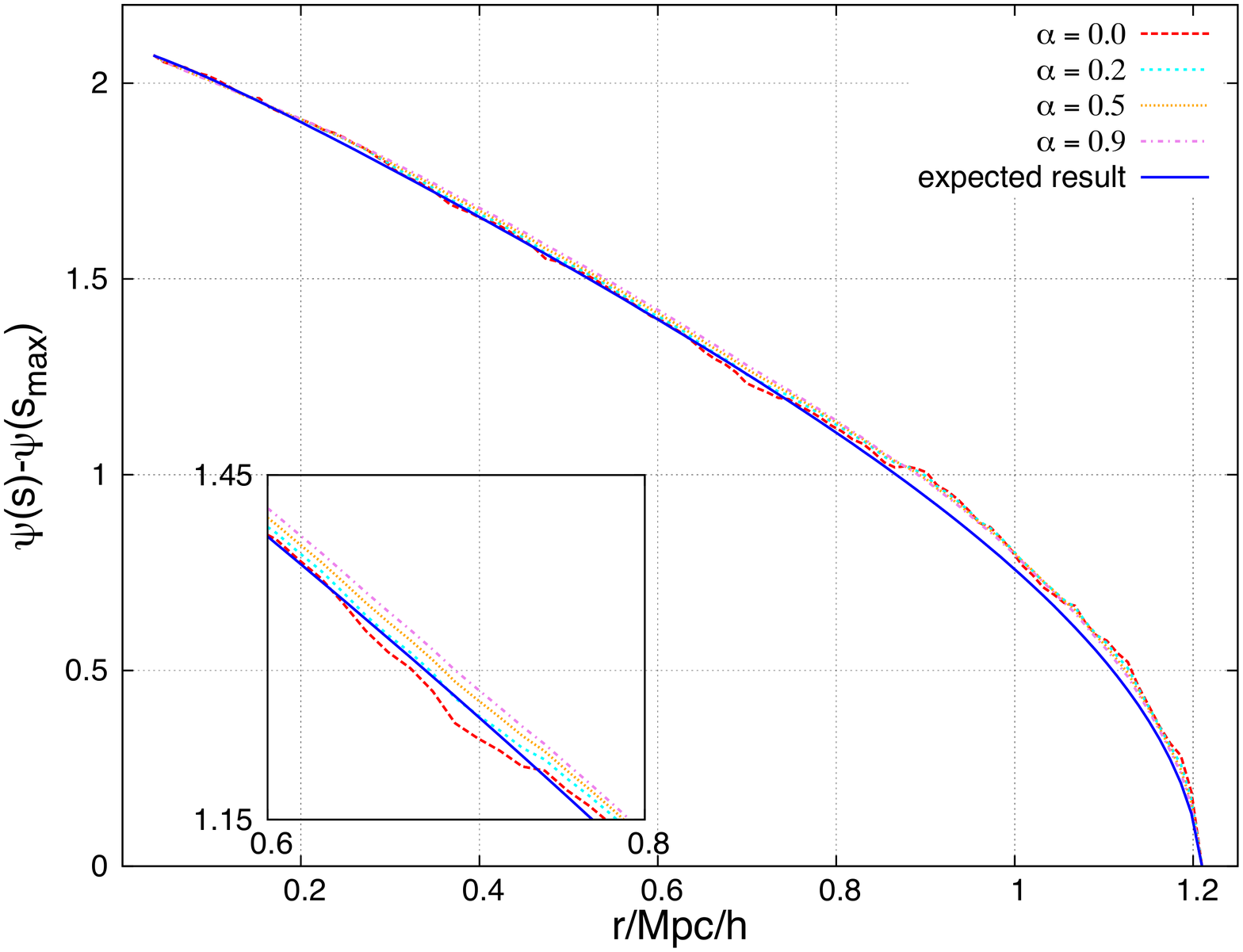}}
\caption{(a) Comparison between potentials recovered with different choices for the amplitude $\alpha$ of the regularisation function in the reconstruction of the effective pressure $P$, keeping the smoothing scale constant at $L=0.3\,h^{-1}\,\mathrm{Mpc}$. The result expected from Eq.~(\ref{eq:6}) is plotted for reference. Varying the smoothing scale within $0.1\le L\le 0.9\,h^{-1}\,\mathrm{Mpc}$ leads to qualitatively similar results. (b) As (a), but for the projected potential $\psi$.}
\label{fig:alpha_L_compare}
\label{fig:alpha_L_compare_psi}
\end{figure*}

\section{Results}

\begin{figure}[!ht]
  \includegraphics[width=\hsize]{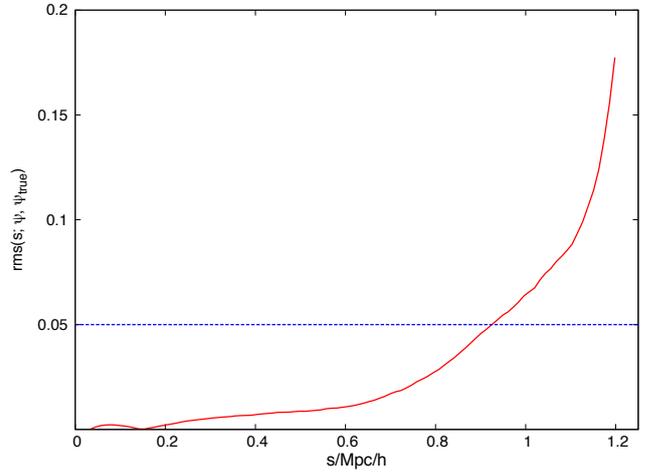}
\caption{Relative \textit{rms} deviation of the projected potential $\psi$ from its true profile $\psi_\mathrm{true}$, derived from 200 realisations of the modelled galaxy cluster with mass $5\times10^{14}\,h^{-1}M_\odot$ at redshift $0.2$. The blue line shows the 5~\% level for comparison.}
\label{fig:residua}
\end{figure}

Applying our algorithm to the simulated specific intensity change $\Delta I_\mathrm{SZ}$ returns an estimate for the three-dimensional effective pressure $P$. Through Eq.~(\ref{eq:6}), an estimate of the gravitational potential $\varphi$ can then be obtained. For the specific data set shown in Fig.~\ref{fig:cluster_profile}, the estimate for the gravitational potential is shown in Fig.~\ref{fig:gravpot}, reconstructed with $\alpha = 0.2$ and $L = 0.3\,h^{-1}\,\mathrm{Mpc}$. As expected, the reconstruction is noisy due to the low signal-to-noise ratio in the input data themselves, even though taking the azimuthal average significantly increases the signal-to-noise ratio. The reconstruction remains reliable to a maximum radius of $r \approx 0.8\,h^{-1}\mathrm{Mpc}$ (cf.\ Fig.~\ref{fig:residua}).

\subsection{Testing the algorithm}

As we have shown in KMMSB, the Richardson-Lucy deprojection algorithm works reliably for reasonable amplitudes $\alpha$ of the regularisation and smoothing scales $L$. We also showed there that an increasing regularisation amplitude can substantially suppress the fluctuations in the reconstructed X-ray surface brightness (see Fig.~\ref{fig:alpha_L_compare_psi}a in KMMSB). Plots showing analogous results for the potential reconstruction from thermal SZ data are shown in Fig.~\ref{fig:alpha_L_compare}a for $0\le\alpha\le0.9$ with a constant $L = 0.3\,h^{-1}\,\mathrm{Mpc}$. Varying the smoothing scale while keeping $\alpha$ constant leads to qualitatively similar results. For large amplitudes $\alpha$, the fluctuations in the recovered potential are damped considerably. For large smoothing scales $L$, the noise is overestimated at large radii.

The effect of varying $\alpha$ and $L$ on the recovered lensing potential is marginal for radii below $r \approx 0.8\,h^{-1}\mathrm{Mpc}$. Only at radii larger than that, the lensing potential becomes smoother and fluctuates less, but tends to overestimate the true lensing potential due to the normalisation conditions of the algorithm. Again, changing $L$ while keeping $\alpha$ fixed has qualitatively similar effects.

\subsection{Error estimation}

We quantify the error of our algorithm by 200 bootstrap resamplings of the galaxy cluster model described above, to which we applied our reconstruction algorithm. From all reconstructions, the rms deviation of the recovered projected potential from its true profile is then calculated,
\begin{equation}
  \mathrm{rms}(s; \psi, \psi_{\mathrm{true}}) = \left[
    \frac{1}{N}\sum_{n=1}^{N}
    \frac
      {\left(\psi^\mathrm{norm}(s)-\psi^\mathrm{norm}_\mathrm{true}(s)\right)^2}
      {{\psi^\mathrm{norm}_\mathrm{true}(s)}^{2}}
  \right]^{1/2}\;,
\label{eq:27}
\end{equation}
where quantities with a superscript `norm' are normalised to reach zero at the maximum projected radius.

The result of this bootstrap is shown in Fig.~\ref{fig:residua} together with a reference line at the 5~\% level of deviation (blue dotted line). We achieve a relative accuracy of less than 5~\% for radii smaller than $\approx 0.9\,h^{-1}\mathrm{Mpc}$. At larger radii, the \textit{rms} reaches values of up to $\lesssim15\,\%$. This is clearly due to the increasing noise in the signal at large radii.

\section{Conclusions}

Following the X-ray analysis presented in the first paper of this series, we have shown in this paper how the observable provided by the thermal SZ effect in clusters, i.e.\ the relative intensity change of the CMB observed through the hot intracluster plasma, can be converted into the projected, two-dimensional gravitational cluster potential. As in the preceding paper, the goal of this study is to bring all cluster observables -- strong and weak gravitational lensing, X-ray emission, the thermal SZ effect and ultimately also galaxy kinematics -- on a common ground to use all of them in a joint reconstruction procedure recovering the gravitational potential best compatible with all these observables.

Assuming hydrostatic equilibrium between the hot gas and the gravitational potential, and further assuming a polytropic gas stratification, we have derived how the Compton-$y$ parameter relates to the gravitational potential. This allowed us to construct an algorithm beginning with the Richardson-Lucy deprojection of the observed, two-dimensional thermal-SZ intensity change into the three-dimensional, effective pressure proportional to the Compton-$y$ parameter. Richardson-Lucy deprojection is the first step in the algorithm requiring symmetry assumptions. For simplicity, not by necessity, we have chosen to assume spherical symmetry for this initial study. The deprojected Compton-$y$ parameter is then readily converted to the three-dimensional gravitational potential, which can finally be projected.

Our implementation of the Richardson-Lucy deprojection algorithm contains an entropic regularisation term with two parameters, an amplitude and a smoothing scale. Both suppress the reconstruction noise as they should, rendering the resulting two-dimensional potential slightly dependent on their values. For quite wide ranges of reasonable parameter choices, however, the result is very close to the expected, two-dimensional cluster potential known from the cluster model underlying the simulations.

We have tested this algorithm with synthetic thermal-SZ data simulated with a spherically-symmetric body of mass $5\times10^{14}\,h^{-1}M_\odot$ at redshift $0.2$, supposed to be observed with the signal-to-noise characteristics of one specific configuration of the ALMA interferometer. In addition to instrumental noise, we have included background fluctuations in the thermal SZ signal due to unresolved clusters. The results look very promising: The three- and two-dimensional gravitational potentials are very well reproduced. Bootstrapping shows relative rms accuracies of the recovered, two-dimensional potential at or below the 5~\% level can be achieved at cluster-centric radii $r\lesssim0.9\,h^{-1}\,\mathrm{Mpc}$.

\acknowledgements{This work was supported in part by the project BA 1369/17 of the Deutsche Forschungsgemeinschaft, by the Collaborative Research Centre TR~33 and by contract research `Internationale Spitzenforschung II-1' of the Baden-W\"urttemberg Stiftung.}

\bibliographystyle{aa}
\bibliography{draft}

\end{document}